\documentclass[english,twocolumn]{revtex4}
\usepackage[T1]{fontenc}
\usepackage[latin9]{inputenc}
\usepackage{graphicx}
\usepackage{amssymb}
\usepackage{amsmath}

\newcommand{\bea}{\begin{eqnarray}}

\newcommand{\eea}{\end{eqnarray}}

\newcommand{\beq}{\begin{equation}}

\newcommand{\eeq}{\end{equation}}

\newcommand{\grad}[1]{\mathbf{\nabla} #1}



\usepackage{babel}

\begin{document}

\title{Realizing and optimizing an atomtronic SQUID}

\author{Amy C. Mathey$^{1}$, and L. Mathey$^{1,2}$}

\affiliation{$^{1}$Zentrum f\"ur Optische Quantentechnologien and Institut f\"ur Laserphysik, Universit\"at Hamburg, 22761 Hamburg, Germany\\
$^{2}$The Hamburg Centre for Ultrafast Imaging, Luruper Chaussee 149, Hamburg 22761, Germany}

\begin{abstract}
 We demonstrate how a toroidal Bose-Einstein condensate with a movable barrier can be used to realize an atomtronic SQUID.  The magnitude of the barrier height, which creates the analogue of an SNS junction, is of crucial importance, as well as its ramp-up and -down protocol. For too low of a barrier, the relaxation  of the system is dynamically suppressed, due to the small rate of phase slips at the barrier. For a higher barrier, the phase coherence across the barrier is suppressed due to thermal fluctuations, which are included in our Truncated Wigner approach. Furthermore, we show that the ramp-up protocol of the barrier can be improved by ramping up its height first, and its velocity after that. This protocol can be further improved by optimizing the ramp-up and ramp-down time scales, 
  which is of direct practical relevance for on-going experimental realizations.   
\end{abstract}
\maketitle
\section{Introduction}
The advancement of  cold atom technology, and the level of control that can be achieved in such systems, has motivated the question if it can be used to emulate electronic circuitry, and possibly move beyond its features, Ref. \cite{holland}. 
 While a realization of, say, the equivalent of electrons moving in a semiconducting material is an interesting direction in itself, it is particularly intriguing to capitalize on the specific features of cold atom systems, such as long-range phase coherence in Bose-Einstein condensates. This motivates to realize systems  inspired by superconducting circuitry.    
 Experimentally, an interesting starting point, and a remarkable achievement of its own, is the realization of Bose-Einstein condensates (BECs) in toroidal geometries, Refs. \cite{gretchensquid, gretchenother, boshier, dalibard}. 
\begin{figure}
   \includegraphics[scale=.48]{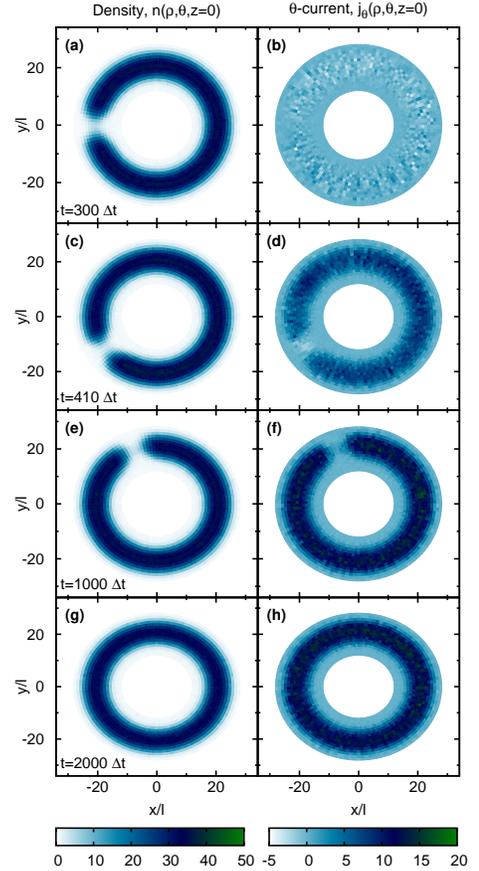} 
    \caption{The density, $n(\rho,\theta)$ and the $\theta$ component of the current, $j_\theta(\rho,\theta)$ at $z=0$ before ($t=300 \Delta t = 217.2$ ms), during ($t=410 \Delta t =296.8$ ms and $t=1000 \Delta t = 724$ ms) and after ($t=2000 \Delta t = 1448$ ms) stirring. The barrier height is $V_b/J=1.8$ and the stirring frequency is $ \omega_s/\omega_0=3.8$.}
    \label{fig:density}
\end{figure}
%
 
 In this paper, we study the equivalent of an electronic SQUID. The condensate wave function is the equivalent of the superconducting wave function, and a potential barrier, at which the condensate density is suppressed,  replaces the SNS interface. This barrier is then moved at a constant speed, which imitates a non-zero magnetic flux through the  ring. This setup can also be seen as a stirring experiment, testing the superfluid properties of  condensates,  see Refs. \cite{matheystirring, stirring}.
  Other theoretical studies of the ring geometry were reported in Refs. \cite{ringtheory}.

 We demonstrate that a regime of  a controlled and effective realization of an atomtronic SQUID exists for sufficiently large potential barrier heights, for realistic temperatures. For small barrier heights, the dynamical relaxation of the condensate to the ground state phase winding is suppressed, because phase slips at the barrier occur only at a very small rate. For larger barrier heights, the phase coherence across the barrier is suppressed because of thermal fluctuations in the bulk of the ring. This results in a smoothed-out response of the rotation, approaching a linear dependence on the stirring velocity, rather than a quantized, step-like response, that is characteristic for a SQUID.  Furthermore, we consider two seemingly similar ramp-up processes for the SQUID operation: to either first ramp up the barrier to full speed and then ramp up the barrier height, or use the reverse order. Interestingly, we find that the latter results in noticeably less heating, and that the new ground state is reached more quickly. We then discuss how this protocol can be further improved by different choices for the ramp-up and -down time scales. 
 
 This paper is organized as follows: In Sect. \ref{method} we describe our simulation method, in Sect. \ref{dynamics} we discuss the phase slip dynamics. In Sect. \ref{optimal} we compare the different barrier ramp-up scenarios, and in Sect. \ref{conclude} we conclude.

\section{Simulation method}\label{method}
Because the SQUID dynamics is dominated by the dynamics at the barrier, and because of the low density at the barrier, and therefore the low mean-field energy, it is imperative to include thermal fluctuations of the system. 
 We include both thermal fluctuations, and the lowest order of quantum fluctuations, within a Truncated Wigner approximation (TWA), see e.g. \cite{TWARevs}. The approach that we use is closely related to the one of Ref. \cite{ring}. 

We describe the system with the Hamiltonian
 \begin{eqnarray}
\hat{H}&=&\int d\mathbf{r} \Big( \hat{\Psi}^\dagger(\mathbf{r}) \left[ -\frac{\hbar^2 \grad^2}{2m} + V(\mathbf{r},t) \right] \hat{\Psi}(\mathbf{r}) \nonumber \\ 
&&+ \frac{g}{2} \hat{\Psi}^\dagger(\mathbf{r}) \hat{\Psi}^\dagger(\mathbf{r}) \hat{\Psi}(\mathbf{r}) \hat{\Psi}(\mathbf{r})\Big)
 \end{eqnarray}
where $m$ is the atomic mass and $g$ is the interaction strength, for which we use the approximation $g=4\pi a_s \hbar^2/m$, with $a_{s}$  being the s-wave scattering length. 
The external potential, $ V(\mathbf{r},t) = V_{tr} (\mathbf{r} )+ V_{bar} (\mathbf{r},t) $ consists of the trapping potential,  
$ V_{tr}(\mathbf{r}) = \frac{1}{2} m \omega_\rho^2(\rho - \rho_0)^2 + \frac{1}{2}m \omega_z^2z^2$, with $\rho=\sqrt{x^{2}+y^{2}}$, and the time-dependent stirring potential, $V_{bar}(\mathbf{r},t) = \alpha_b(t)V_b\exp\left[- 0.5 \rho^2(\theta-\theta_b(t))^2/l_b^2\right]$, with $\theta$ being the azimuthal angle. $\rho_{0}$ is the radius to the ring.  
 Within the TWA, the operators, $\hat{\Psi}$ are replaced by classical fields, which are propagated according to the equations of motion of this    Hamiltonian. The initial condition of these classical fields are generated from the Wigner distribution of the initial state. 

In order to carry out the calculations, we discretize the real-space description by introducing a lattice approximation and work in cylindrical coordinates.   We replace the continuous wave function, $\psi(\rho,\theta,z)$ by a discrete wavefunction, $\tilde{\psi}_{ijk}=\tilde{\psi}(\rho_i,\theta_j,z_k) $, with the mapping
$$
\psi(\rho,\theta,z)  \rightarrow \left( \frac{\rho_0}{\rho_i l^3}\right)^{1/2}\tilde{\psi}_{ijk}
$$
 where $\rho_i=\rho_0 + l\left[ i -(N_\rho-1)/2 \right]$, $i \in [ 0, ..., N_\rho-1]$, 
 $\theta_j=l j/\rho_0$,  $j \in [0,...,N_\theta-1] $, and  $z_k = \left[k - (N_z-1)/2\right]l$, $k \in [0,...,N_z-1]$, and $\rho_0=lN_\theta/(2\pi)$ is the radius of the ring at the trap minimum. 
  We emphasize that the discretization length is not constant, but increases linearly with increasing distance from the center axis of the ring. Therefore, the curvature of the ring geometry is fully taken into account.
  In this representation, $|\tilde{\psi}_{ijk}|^2$ corresponds to the number of atoms per unit cell, where the volume of the unit cell is $l^3 \rho_i/\rho_0$.  For the lattice size chosen here, the volume of the unit cell varies from $0.6 l^3$ for $i=0$ to $1.4l^3$ for $i=16$.

 In this representation, the equations of motion take the form
 \begin{eqnarray} 
 i\hbar \frac{d}{dt}\tilde{\psi}_{ijk} &=& -J \Bigg[ \left( \tilde{\psi}_{i+1jk} +\tilde{\psi}_{i-1jk} - 2\tilde{\psi}_{ijk}  \right)
 +\frac{l^2}{4\rho_i^2}\tilde{\psi}_{ijk}\nonumber \\
&&  + \frac{\rho_0^2}{\rho_i^2}\left(\tilde{\psi}_{ij+1k}  +\tilde{\psi}_{ij-1k}-2\tilde{\psi}_{ijk} \right)\nonumber \\
&&  + \left(\tilde{\psi}_{ijk+1}+\tilde{\psi}_{ijk-1}-2\tilde{\psi}_{ijk} \right) \Bigg]\nonumber \\
&&+\left[  V_{ijk} +U(\rho_i)|\tilde{\psi}_{ijk} |^2\right] \tilde{\psi}_{ijk}, \label{eqn:eom}
\end{eqnarray}
where $J=\hbar^2/(2ml^2)$ is the tunneling energy, and the interaction term is given by $U(\rho_i) =  U_0 \rho_0/\rho_i$, where $U_0= gl^{-3}$.  
 On the lattice, the external potential is given by $V_{ijk} = \frac{\hbar^2}{4Jl^2} \left[\omega_\rho^2\left(\rho_i - \rho_0\right)^2 + \omega_z^2 z_k^2 \right]  +\alpha(t)V_b\exp\left[- 0.5 \rho_i^2(\theta_j-\theta_b(t))^2/l_b^2\right]$. 
 
 As mentioned above, we initialize the dynamics by sampling from the initial Wigner distribution. For the initial state, we choose a   non-interacting Bose gas in the toroidal trap, of zero temperature.  After the initialization, the interactions are turned on slowly to generate the desired  interacting ensemble in the trap \footnote{The eigenstates of the non-interacting Bose gas are generated by diagonalizing the non-interacting Hamiltonian on the lattice in $\rho$ and $z$ directions, and using plane waves along $\theta$. The interaction strength is turned on slowly, according to $\alpha_U(t)= (1+\tanh\left\lbrace 5[(t-100)/8000-0.5]\right \rbrace )/2$. The total initialization time is $16000 \Delta t$}.
  This process of turning on the interaction results in a non-zero temperature that is comparable or larger than the mean-field energy of the system
   \footnote{Because the temperature is comparable or larger than the mean-field energy, the thermal fluctuations of the system dominate over the quantum fluctuations. We could therefore also use  Monte Carlo sampling of the initial state as well, as it was done in Ref. \cite{matheystirring}. The method we use in this paper was merely chosen for its efficiency for the system at hand. }. 
 This  temperature  is measured by weakly coupling harmonic oscillators to the current as described in Ref. \cite{ring}. For examples discussed in this paper, the temperature of the atomic cloud after initialization is approximately  4.0 J = 43 nK \footnote{Using the notation of \cite{ring}, the harmonic oscillator parameters are $\omega_{ho}/J = 1.0$, $U_{ho}/J=0.02$ and the coupling is turned on and off over a time scale $\tau_{ho}/J=4000$.}.

 Throughout this paper, we use a lattice with the dimensions $N_{\rho}=17$, $N_{\theta}=126$, and $N_z=5$, and $N=50000$ atoms. The trapping frequencies are $\omega_\rho/J = 0.5$ and $\omega_z/J = 2.5$.   We propagate and average over 72 initial states and the length scale of the barrier is $l_b/l=3$. 
  We set $U_0/J=0.07$, which corresponds to a length scale $l=0.99 \mu$m, a time scale $\Delta t = \hbar/J = 0.71$ ms and an energy scale $J=10.8$ nK for sodium atoms.
 The healing length in the bulk of the system is $\xi \approx 0.9 \mu$m. This length is small compared to system size in the radial direction and comparable to the system size in the $z$-direction, which puts the system in the dimensional cross-over regime between two and three dimensions.

\begin{figure}
   \includegraphics[scale=.58]{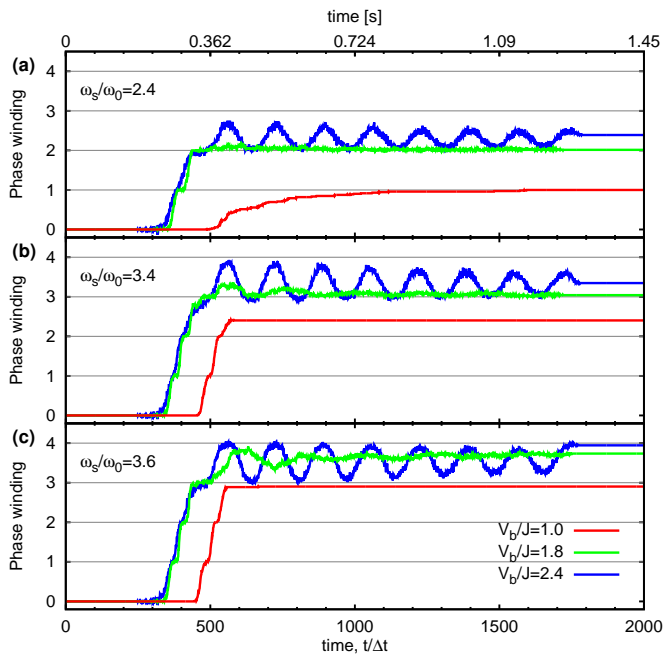} 
   \caption{We depict the time evolution of the average phase winding around the center of the toroid for stirring frequencies, $\omega_s/\omega_0= 2.6, 3.4, 3.6$, and barrier heights $V_b/J=1.0, 1.8, 2.4$.}
   \label{fig:phwdtime}
\end{figure}

We now consider the following experiment:
Starting at $t_1=100 \Delta t$, we ramp the barrier magnitude from 0 to the final barrier height $V_{b}$ over  a time period of $200\Delta t$, while keeping the barrier stationary. Then at $t_2= 300 \Delta t$, the barrier is accelerated to its final stirring frequency, $\omega_s$ over $200 \Delta t$. At $t_3=500\Delta t$, the atomic cloud is stirred at constant frequency for a time period of $1200\Delta t$ at maximum barrier height.  At $t_4=1700\Delta t$ the barrier height is ramped down over the time $200\Delta t$ while continuing to stir the atoms at the frequency $\omega_s$.  We refer to this procedure as protocol 1.

 Additionally, we compare this protocol to the process of turning on the barrier height to $V_{b}$, while stirring at constant frequency $\omega_s$, starting at $t_1$ and reaching the maximum barrier height at $t_2$.  The atomic cloud is stirred for $1400\Delta t$ before ramping down the barrier, while continuing to stir at constant frequency. This second protocol is reminiscent of the experiments reported in Ref. \cite{gretchensquid}.  As we discuss below, the first protocol is the preferable protocol for large stirring frequencies, because it avoids exciting phase slips before the barrier has reached its maximum height.
  We elaborate on this protocol further in Sect. \ref{optimal} by considering different ramping time scales.

  In Fig. \ref{fig:density}, we illustrate the dynamics of the system by depicting the density $n(\rho,\theta)$ in the $z=0$ layer, and the current $j_\theta(\rho,\theta)$ along the azimuthal direction of the ring, at four different times, for a barrier height of $V_b/J=1.8$ and stirring frequency $\omega_s/\omega_0=3.8$ for the first protocol, with $\omega_{0} = \hbar/(m \rho_{0}^{2})$.
  At the first time, $t = 300 \Delta t$, the trap potential is fully ramped up, as is visible in the density depletion of the ring shaped condensate, but is still stationary. Here, the phase winding is still zero, and both positive and negative current fluctuations are visible. We note again that these fluctuations are predominantly of thermal origin.
     At the second time, $t = 410 \Delta t$, one phase slip has occurred. Now the current has acquired a preferred direction, as is immediately visible in Fig. \ref{fig:density} (d).
       At the third time, $t = 1000 \Delta t$, a second phase slip has occurred, and the magnitude of the current has increased, Fig. \ref{fig:density} (f).
     Finally, at time $t = 2000 \Delta t$, the barrier has been ramped down, so the density of the condensate does not display a depleted region. However, as is visible from the current in  Fig. \ref{fig:density} (h), the stirring of this ring shaped condensate has imparted a finite current circulating in the ring.

For a system with a complex order parameter, such as a condensate of atoms or a superconductor, this current is related to the well-defined phase. In the condensed phase, this results in a quantization of the phase winding around a ring geometry. 
 We determine the phase along the central line along the ring, which tracks the maximal density for a given azimuthal angle, and in the $z=0$ plane.
   We calculate the phase at the angle $\theta$ via 
 \bea
 \phi(\mathbf{\theta} )= \phi(\rho_0,\theta,0) & = & \tan^{-1}(\text{Im }\psi (\mathbf{\theta})/\text{Re } \psi(\mathbf{\theta})).
 \eea
  The phase winding is the sum of the phase differences around the ring,
$n_{\Delta \phi} =\left( \sum_\theta \delta \phi(\mathbf{\theta}) \right) /2\pi $,  where the phase difference between two points, $\delta \phi (\theta) = \phi(\mathbf{\theta+\theta_0}) - \phi(\mathbf{\theta}) $ is between $ -\pi$ and $\pi$.
The phase winding is calculated in each individual realization and then averaged over the realizations to generate the average phase winding.

\begin{figure}
   \includegraphics[scale=.58]{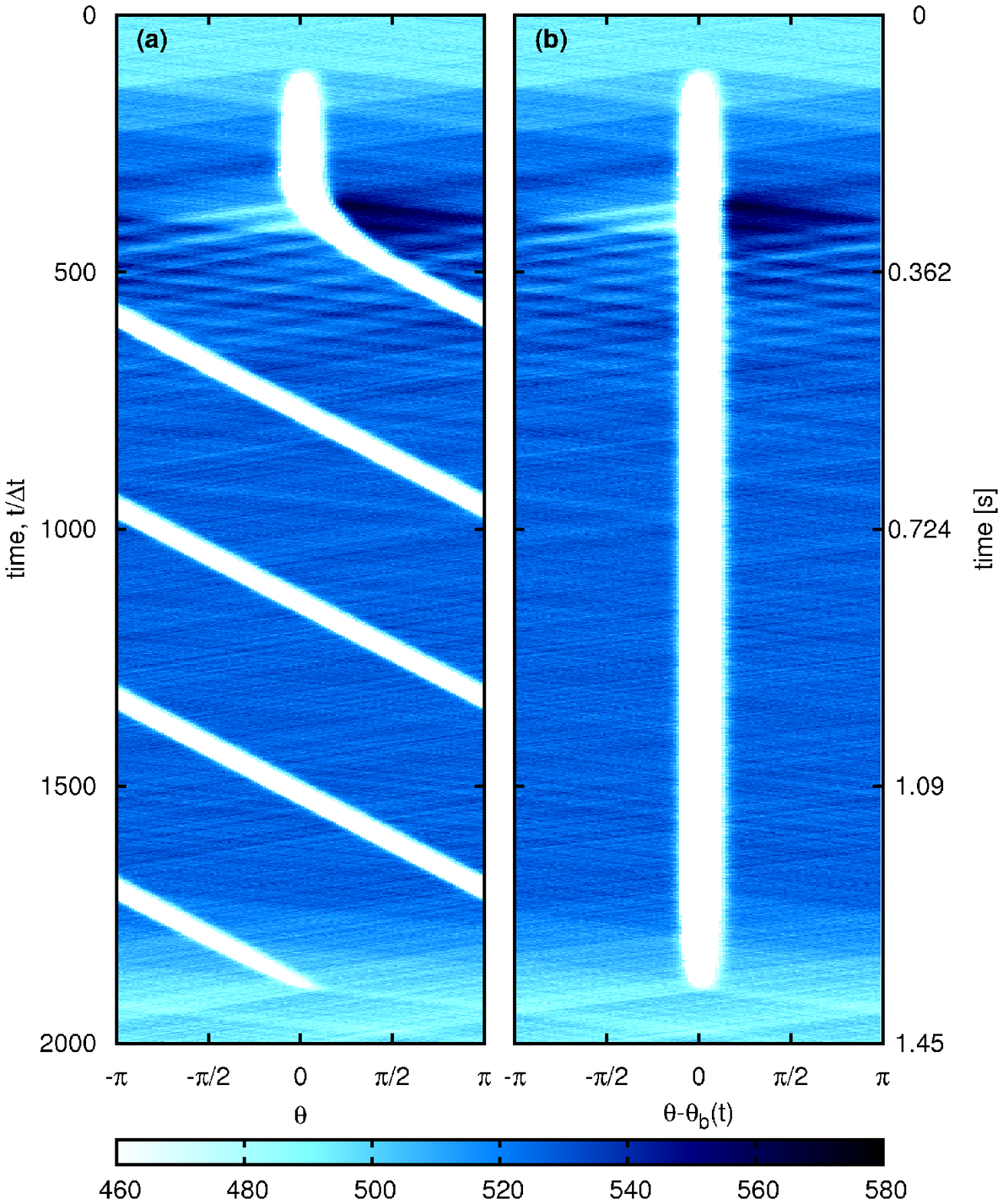} 
   \includegraphics[scale=.58]{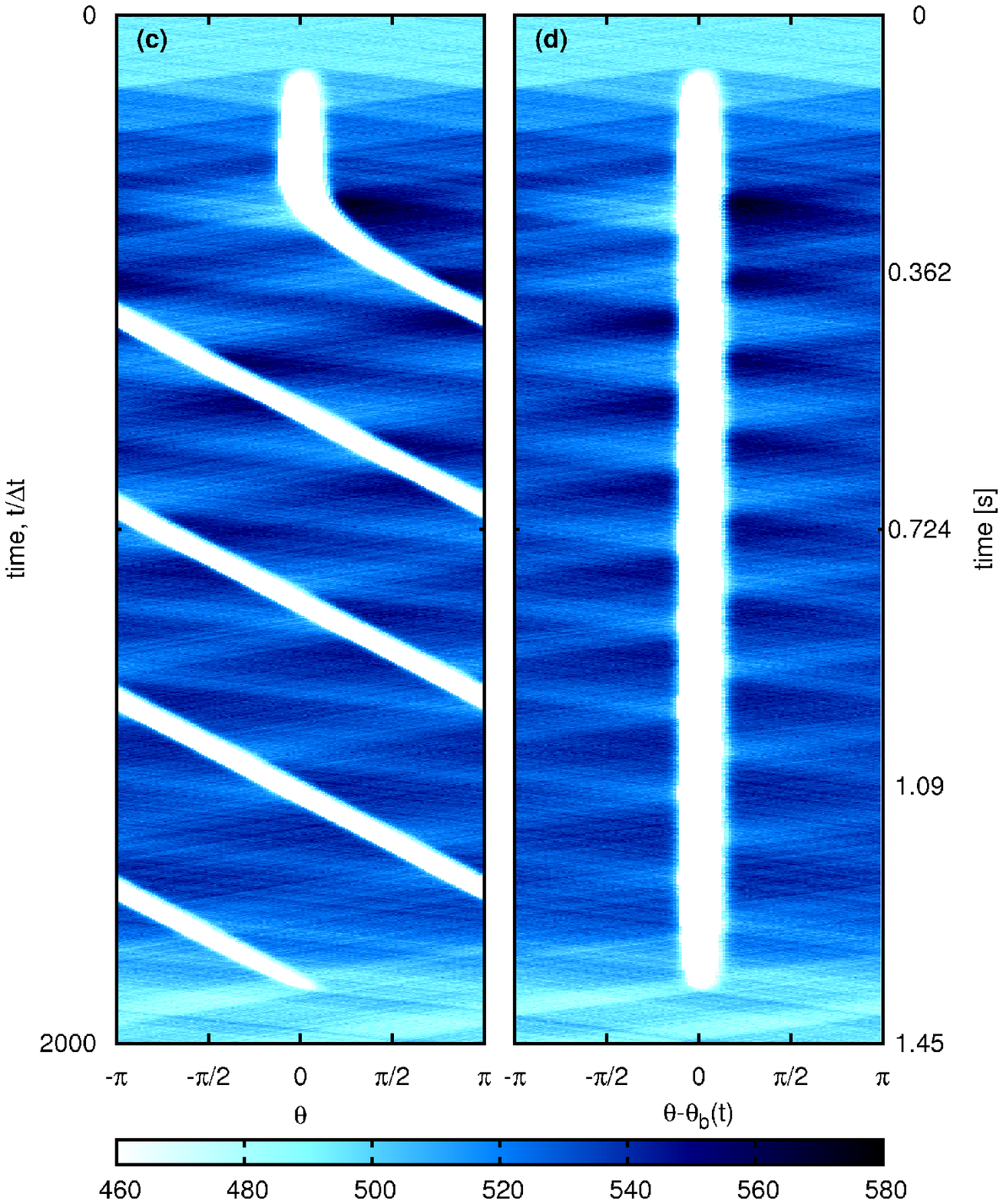} 
   \caption{We depict the time evolution of the radially projected density, $n_{1D}(\theta)$,  for $\omega_s/\omega_0=3.4$. In panels (a) and (b), we use the barrier height $V_{b}/J=1.8$, and (c ) and (d) $V_b/J=2.4$.  In (b) and (d) the density is plotted in the reference frame of the barrier. The phonons travel as straight lines in this depiction.}
   \label{fig:1Ddensity}
\end{figure}
\section{Phase Slip Dynamics}\label{dynamics}
In Fig. \ref{fig:phwdtime} we show the time evolution of the average phase winding along the center of the toroidal trap for three different barrier heights and three different stirring frequencies.  For the smallest barrier height of $V_{b}/J = 1$, and for all three rotation frequencies $\omega_{s}$, the phase winding does not relax to the new ground state. Instead, on the time scales of a typical experiment, it remains in a long-lived, metastable state. 
 On the other hand, for large barrier heights such as $V_{b}/J = 2.4$, the coupling across the barrier is  small compared to the temperature, so that the dynamics are oscillatory and noisy for the length of the experiment. Here, the dynamics of the experiment results in a distribution of phase windings, rather than - essentially - a single phase winding number.  
  However, we observe that for intermediate barrier heights such as $V_{b}/J = 1.8$, the oscillations damp out in a realistic time.

We note that the magnitude the phase slip rate, and its dependence on the system properties war discussed in Ref. \cite{ring}. In particular, it was discussed that the phase slip rate is consistent with the scaling $\tau^{-1}_{ph} \sim \exp(- E_{b}/k_{B} T)$, i.e. an Arrhenius law. The energy scale $E_{b}$ is controlled by the barrier height and width.
 This strong, exponential dependence is reflected in the behavior we observe here, where the relaxation dynamics of the system is nearly suppressed at $V_{b}/J = 1.0$, while at around $V_{b}/J \approx 2.4$ the barrier is so high that any coherence across the barrier is suppressed.

 To understand the origin of the oscillatory behavior we show the time evolution of the density along the ring. We define radially projected density 
  $n_{1D}(\theta)=\int{d\rho} \int{dz} n(\rho,\theta,z)$. 
   We emphasize that while this quantity has the dimensions of a one-dimensional density we do not imply that the dynamics can be reduced
    to that of a one-dimensional system. In fact, as we had discussed in Ref. \cite{ring}, the phase slips that occur in the system are due to a non-trivial process of a vortex traversing the barrier region. We merely use this integrated density as a convenient way to depict the system evolution. 
    
 In Figure \ref{fig:1Ddensity} (a) and (c ) the time evolution of the radially projected density is shown for $V_b/J=1.8$ and $V_b/J=2.4$, respectively,  and for $\omega_s/\omega_0=3.4$.   In (b) and (d), the same evolution  is shown in the reference frame of the barrier.
  As is visible in this figure, 
  the oscillations in the average phase winding are due phonon pulses generated during the initialization and acceleration of the barrier.    The density waves observed in the  density travel at the speed of sound, which is approximately $v_s=1.52 l/\Delta t$ or $2.1$mm/s.  With this velocity, the period of the oscillations 
  of the average phase winding is of the order of $2T=166 \Delta t$, corresponding to traveling back and forth along the circumference of the ring. 
  This  matches the period of the oscillations observed in Fig. \ref{fig:phwdtime}.
   We observe that the oscillations for  $V_b/J=1.8$ damp out during the time of the simulation, whereas for  $V_b/J=2.4$ they do not.
    This is due to the reflection of the phonon pulse at the barrier. For the high barrier this reflection is essentially complete, and the pulse travels back and forth with only little damping. For the intermediate barrier height, the reflection is partial, which results in dephasing and damping.
  
   Motivated by this observation, 
     we suggest ways to minimize the detrimental effect of these phonons pulses and the resulting oscillations in the phase winding in Sect. \ref{optimal}, as a step towards improving the SQUID operation of the condensate ring.

\begin{figure}
   \includegraphics[scale=.58]{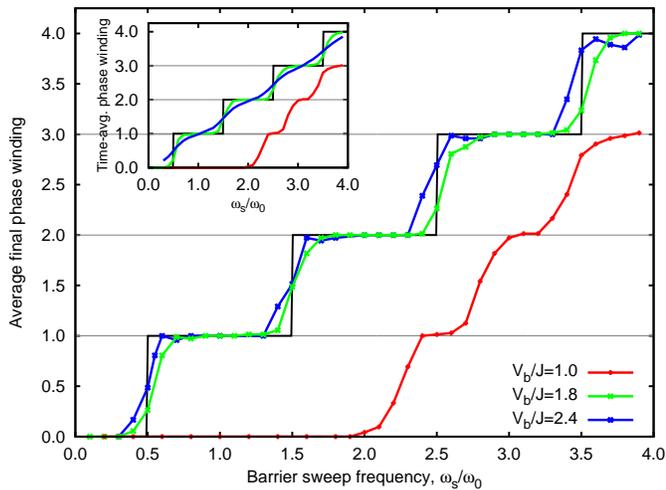} 
    \caption{We depict the average final phase winding as a function of the barrier sweep frequency for barrier heights, $V_b/J=1.0, 1.8, 2.4$. For $V_{b}/J =1$, the system is remains dynamically trapped below the equilibrium phase winding number. For  $V_{b}/J = 1.8$ and $V_{b}/J = 2.4$ a step-like response is achieved. The black line indicates steps at $\omega_s/\omega_0 = n+1/2$, where $n$ is an integer.   
     The inset shows the time averaged phase winding before the barrier is ramped back down.
    For large barrier heights the system approaches a linear response. This tendency is already visible for $V_{b}/J = 2.4$, and continues for higher values.
    }
    \label{fig:steps}
\end{figure}

The phase winding that emerges from this dynamical evolution is shown in Fig. \ref{fig:steps}. 
 We depict the final, average phase winding after the barrier has been ramped down, as a function of the stirring frequency for several barrier heights, in the main panel. 
   As indicated, the idealized, fully quantized, and fully relaxed phase winding shows steps at $\omega_{s}/\omega_{0} = n + 1/2$, with $n$ being an integer. This is shown as a black line.
        For too small of a barrier height, such as $V_{b}/J= 1$, the system remains dynamically trapped at a smaller phase winding than for the fully relaxed system.
   As the barrier height is increased, the behavior of the phase winding approaches a smoothed-out step-like behavior, for   $V_{b}/J= 1.8$ and  $V_{b}/J= 2.4$. 
    We note that this smoothed out step-like response in the  phase winding is comparable to the results that were obtained experimentally in Refs. \cite{gretchensquid}. 
       In the inset of Fig.~\ref{fig:steps}, we plot the average phase winding, time-averaged over four phonon periods, $[t_4-4T,t_4]$, before the barrier is ramped down. 
This is the phase winding that the system relaxes to at longer times, after the oscillatory behavior has damped down, with the stirring on.  
 We note that for $V_b/J=1.0$ and $V_b/J=1.8$ the time-averaged phase winding is very close to the phase winding shown in the main panel, because the oscillatory behavior has damped out on the time scale of the experiment.  
  However, for $V_b/J=2.4$, the phase winding that is shown in the inset is smoothed out to an almost linear behavior. This indicates that 
 the classical limit of this response is nearly reached, as expected for a fully disconnected ring. 
 We therefore conclude  that this limit should be observed in experiment for high barrier potentials, at long stirring times and instantaneous ramp-down of the barrier.
  Furthermore, we conclude that the re-emergence of the step-like behavior after the ramp-down of the barrier, is due to the non-zero time of this ramp-down, during which the system develops a well-defined phase winding. This motivates our proposal to increase this ramp-down time in Sect. \ref{optimal}.

\begin{figure}
   \includegraphics[scale=.58]{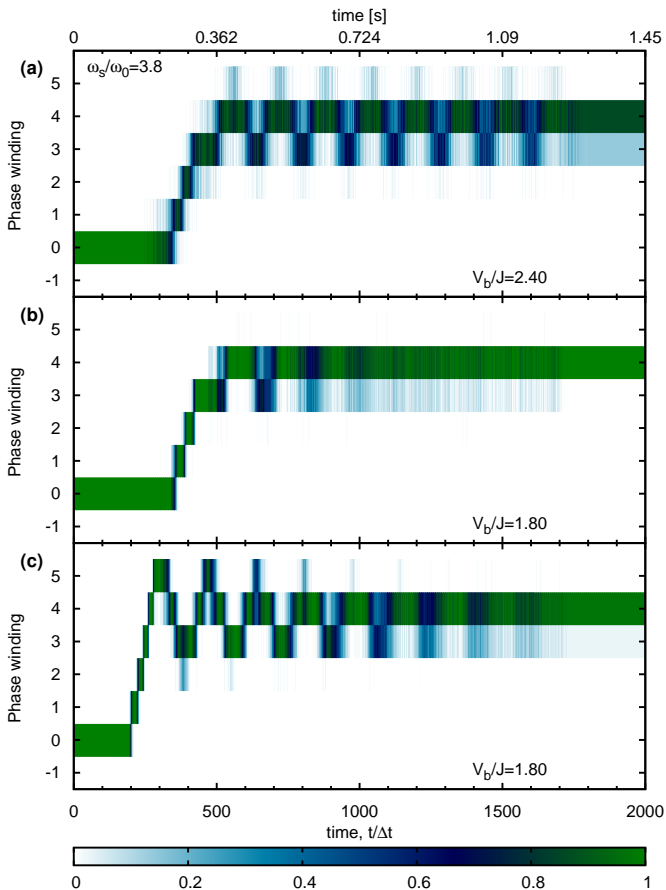} 
    \caption{We plot the occupation of each phase winding for sweep frequency  $\omega_s/\omega_0=3.8$ and barrier heights $V_b/J=1.8, 2.4$. In (a) and (b), the barrier is stationary while it is ramped up, from $t_1=100 \Delta t$ to $t_2=300 \Delta t$ and then accelerated at maximum height, from $t_2=300 \Delta t$ to $t_3=500 \Delta t$ (protocol 1). In (c), the barrier stirs with a constant frequency while the magnitude is ramped up, from $t_1=100 \Delta t$ to $t_2=300 \Delta t$ (protocol 2).}
    \label{fig:histogram}
\end{figure}

 In Fig. \ref{fig:histogram} (a) and (b) we elaborate on the behavior shown in Fig. \ref{fig:steps}. We show the distribution of the phase winding, as a function of time, which goes beyond the expectation value of this distribution that was shown in Fig. \ref{fig:steps}.   In Fig. \ref{fig:histogram} (a) we show the case of the larger value of the barrier height, $V_{b}/J = 2.4$, and in Fig. \ref{fig:histogram} (b) we show the case with $V_{b}/J = 1.8$. For the intermediate barrier height of $V_{b}/J = 1.8$, the system converges to a single value of the phase winding over the time of the experiment. The transient oscillations are damped out on this time scale. For the larger value of $V_{b}/J = 2.4$ the distribution of phase windings is wider and stays oscillatory throughout the experiment time. It does not settle to a single value, but rather a distribution that mostly includes $n_{\Delta \theta} = 3$ and $4$, in this example. 
 
 We note that the width of this distribution is controlled by the long range phase fluctuations along the quasi-1D geometry of the ring-shaped condensate. As discussed in Ref. \cite{phasefluct}, the single particle correlation function along the ring falls off exponentially, with a length scale $l_{\phi} = \hbar^{2} N_{0}/(\pi m \rho_{0} k_{B} T)$. $N_{0}$ is the number of condensed atoms. 
 This length scale has to be compared to the circumference of the ring, $L_{c} =2 \pi \rho_{0} \approx 124.4 \mu$m.
  For $N_{0}\approx N$, and $N$ being the total atom number, and for $T = 50$ nK, we have $l_{\phi} = 336 \mu$m. This results in a ratio $L_{c}/l_{\phi}$ that is smaller than $1$. However, this does suggest, that the phase coherence  across the barrier can be further stabilized by increasing this ratio. 
 This ratio can also be written as
 \bea
 \frac{L_{c}}{ l_{\phi}} &=& \frac{4 \pi^{2}}{ n_{2D} \lambda_{T}^{2}}
 \eea
 where $\lambda_{T}$ is the thermal de Broglie wavelength, $\lambda_{T} = (2\pi \hbar^{2} /(m k_{B} T))^{1/2}$, and $n_{2D} = N/ (\pi \rho_{0}^{2})$ is a the density of a hypothetical system of area $\pi \rho_{0}^{2}$ with $N$ atoms. 
  Another way of stating the meaning of this ratio is that it describes the magnitude of the phase difference across the barrier, $(\Delta \phi)^{2} \sim L_{c}/l_{\phi}$. 
 Therefore, the figure of merit that determines if the  elongated 3D condensate is phase coherent along its extended 1D axis, is of the form of an inverse 2D phase space density.
  The optimal regime is that of low temperatures, and ring condensate with a small radius and  high density. 
  If, on the other hand, one wants to explore the effective 1D regime of a phase-fluctuating condensate, this figure of merit has to be increased above $1$. 
 

\section{Optimizing the barrier protocol}\label{optimal}
 \begin{figure}
   \includegraphics[scale=.58]{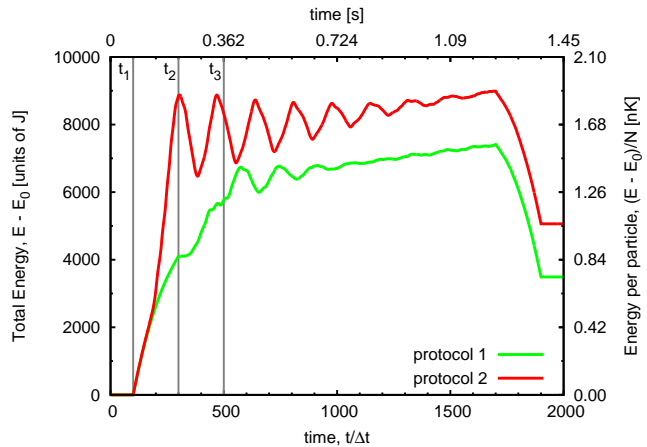} 
    \caption{We depict the dynamics of the energy for sweep frequency $\omega_s/\omega_0=3.8$ and barrier height, $V_b/J=1.8$. In protocol 1 the barrier is stationary while it is ramped up, from $t_1=100 \Delta t$ to $t_2=300\Delta t$ and then accelerated at maximum height, from $t_2=300 \Delta t$ to $t_3=500 \Delta t$. In protocol 2, the barrier stirs at constant frequency while the magnitude is ramped up, from $t_1=100 \Delta t$ to $t_2=300 \Delta t$. $E_0$ is the total energy at t=0.}
    \label{fig:heating}
\end{figure}
   As a last point, we compare the two ramp-up protocols, mentioned above, as well as different ramping times for the barrier. 
        In  Fig. \ref{fig:histogram} (b) we show the case in which the barrier height is ramped up first, while remaining stationary, and then its velocity. This is the case which has been discussed thoughout this paper.
 In Fig. \ref{fig:histogram} (c) we show the case in which the velocity is always at its final magnitude, and the height is ramped up from zero. 
  We see that in the latter case the response is visibly more oscillatory and noisy. Here, the phase slips begin to occurs around $t=200 \Delta t$, when the barrier is still well below its maximum height.  Each phase slip generates phononic excitations which are released into the bulk of the condensate, as visible in Fig.~\ref{fig:histogram} (c).  Similar processes were observed in Ref. \cite{ring}.  Additionally, we observe that it takes longer for the system to relax to a single phase winding. This suggests that the density at the barrier should be minimized when the phase slips occur to reduce undesirable excitations. 
    
This is illustrated in Fig. \ref{fig:heating}. Here, we show the total energy per particle during the evolution of these two protocols. A larger magnitude of this quantity will result in a higher temperature of the system after it has thermalized, and therefore can be used as a measure for how well the SQUID is implemented.   As visible, in this protocol additional undesired excitations are created, along with the desired phase slips, which leads to longer relaxation times and additional heating of the system.  These effects are more pronounced at higher stirring frequencies than at lower stirring frequencies and are expected to play a bigger role as the stirring frequency is further increased.  Again, we observe that the preferable operation of the SQUID consists of first ramping up the barrier height, and then the barrier velocity.

\begin{figure}
   \includegraphics[scale=.58]{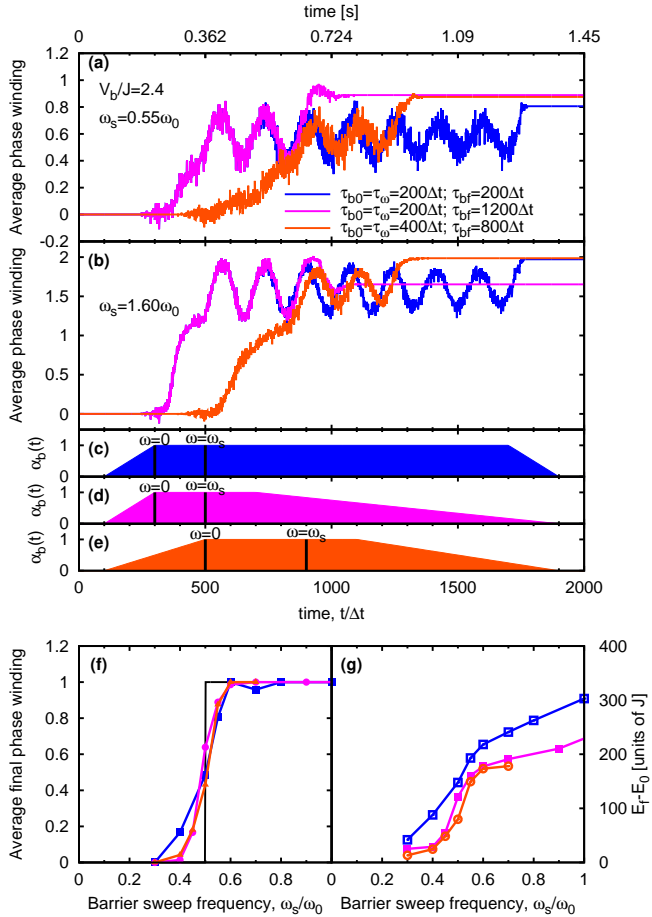} 
    \caption{(a) Time dependence of the average phase winding for three different barrier sequences.  
Starting at $t_1=100 \Delta t$, the barrier height is ramped up over $\tau_{b0}$, then accelerated over $\tau_\omega$.  The cloud is stirred at constant angular velocity and the barrier is ramped down over $\tau_{bf}$, starting at $1900\Delta t-\tau_{bf}$.  The normalized barrier height is depicted in (c ) $\tau_{b0}= 200 \Delta t$,  $\tau_{\omega}= 200 \Delta t$, $\tau_{bf}= 200 \Delta t$, (d)  $\tau_{b0}= 200 \Delta t$,  $\tau_{\omega}= 200 \Delta t$, $\tau_{bf}= 1200 \Delta t$,  and (e) $\tau_{b0}= 400 \Delta t$,  $\tau_{\omega}= 400 \Delta t$, $\tau_{bf}= 800 \Delta t$. Average final phase winding (f)  and final energy per particle (g) as a function of the stirring frequency.  The maximum barrier height is $V_b/J=2.4$ and the stirring frequency is $\omega_s=0.55 \omega_0$ for (a) and (c )--(g). In (b) we show the phase winding obtained for $\omega_{s}/\omega_{0} = 1.6$. }
    \label{fig:optimal}
\end{figure}

 Next we address the influence of the ramp times on the SQUID operation. 
  As a key example, we focus on the first step of the phase winding, around $\omega_s/\omega_0 \approx 1/2$. 
   In Fig.~\ref{fig:optimal}, we consider three different time sequences.
    The sequence that we discussed up to here, is shown in panel (c ). Additionally, we consider two other sequences, shown in (d) and (e).
   The sequence in (d) features a slow ramp down, and the sequence in (e) features both a slow ramp-up and a slow ramp down.
   For stirring frequencies near the first step, these two sequences both lead to an improvement: In panel (f) we show the resulting phase winding. The phase winding increases more steeply for the time sequences with slower ramp-up and ramp-down. 
    Furthermore, in panel (g) we show the resulting increase of the energy of the system, which is again improved for the slower ramp times.
    For these comparatively small stirring frequencies, the improvement is primarily due to the slower ramp-down.
    For higher stirring frequencies, the slower ramp-up time leads to a further improvement of the operation, because the phonon pulse that is created by the barrier acceleration is reduced. We give an indication for this behavior in panel (b). Among the three phase winding evolutions that are shown, the protocol that is shown in panel (e) is the least oscillatory.  
       
       Furthermore, we point out that the slow ramp-down time also allows for the sloshing motion of the system to damp down. The detrimental feature of this motion is not due to the magnitude of the phase fluctuations, but because it can the lead the system to arrive at a phase winding other than the ground state one, if the barrier is ramped down fast.
        As mentioned above, this phonon motion damps out for intermediate values of the barrier, but not for high barriers. If the ramp-down is done sufficiently slow, this provides  enough time for the oscillations to damp out, while the barrier is at intermediate values.

\section{Conclusions}\label{conclude}
In conclusion, we  have demonstrated the atomtronic implementation of a SQUID in a toroidal Bose-Einstein condensate, with particular emphasis on the effect of thermal fluctuations. These are included in our Truncated Wigner approximation of a realistic system, as realized in the experiments of Refs. \cite{gretchensquid, gretchenother}.
 We show that the regime of SQUID operation is viable for sufficiently large barrier heights, which imitates an SNS junction of a solid state SQUID. For too low of a barrier height, the rate of phase slips is too low for the system to reach the equilibrium  phase winding number for a given moving barrier speed. 
  For larger barrier heights, the thermal phase fluctuations suppress the coherence across the barrier. The characteristic, step-like behavior of the phase winding is achieved  
   for either intermediate values of the barrier height, or during the ramp-down of the barrier, if this occurs on a sufficiently long time scale.
   Furthermore, we investigate two ramp-up protocols of the barrier height and the barrier velocity, and show that ramping up the barrier height first, before setting it in motion,  results in less heating.
   An additional improvement can be achieved by increasing both the ramp-up time of the barrier and the ramp-down time. The slower ramp-up results in a reduction of the phonon pulse that is emitted when the barrier is set in motion. The slower ramp-down of the barrier improves the reemergence of an integer phase winding after the stirring. 
 We emphasize that these results and considerations will similarly apply to all atomtronic circuits, that are based on condensate dynamics in non-trivial trap geometries, and are therefore of broad interest to the emerging field of imitating superconducting circuitry with Bose-Einstein condensates.

\begin{acknowledgments}
 We acknowledge support from the Deutsche Forschungsgemeinschaft through the SFB 925 and the Hamburg Centre for Ultrafast Imaging, and from the Landesexzellenzinitiative Hamburg, supported by the Joachim Herz Stiftung.

\end{acknowledgments}

\end{document}